\documentclass{svjour2}
\usepackage{graphicx}

\journalname{General Relativity and Gravitation}

\def\beq{\begin{equation}}
\def\eeq{\end{equation}}

\def\rmd{{\rm d}}

\begin{document}

\title{Dixon's extended bodies and weak gravitational waves}

\author{Donato Bini \and
        Christian Cherubini \and
        Andrea Geralico  \and
        Antonello Ortolan
}

\institute{Donato Bini 
              \at
              Istituto per le Applicazioni del Calcolo ``M. Picone,'' CNR, I-00161 Rome, Italy\\
              ICRA, University of Rome ``La Sapienza,'' I--00185 Rome, Italy\\
              INFN - Sezione di Firenze, Polo Scientifico, Via Sansone 1, I--50019, Sesto Fiorentino (FI), Italy\\
              \email{binid@icra.it} 
    \and
              Christian Cherubini 
              \at
              Facolt\`a di Ingegneria, Universit\`a Campus Biomedico, Via Alvaro del Portillo 21,  I-00128 Roma, Italy\\
              ICRA, University of Rome ``La Sapienza,'' I--00185 Rome, Italy\\
              \email{cherubini@icra.it}         
    \and
          Andrea Geralico 
              \at
              Physics Department and ICRA, University of Rome ``La Sapienza,'' I--00185 Rome, Italy\\
              \email{geralico@icra.it}
    \and
          Antonello Ortolan 
              \at
              INFN - National Laboratories of Legnaro, I-35020 Legnaro (PD), Italy\\
              \email{ortolan@lnl.infn.it}
}

\date{Received: date / Accepted: date / Version: date}

\maketitle

\begin{abstract}
General relativity considers Dixon's theory as the standard theory to deal with the motion of extended bodies in a given gravitational background.
We discuss here the features of the \lq\lq reaction'' of an extended body to the passage of a weak gravitational wave.
We find that the  body acquires a dipolar moment induced by its quadrupole structure.
Furthermore, we derive the \lq\lq world function'' for the weak field limit of a gravitational wave background and use it to estimate the deviation between geodesics and the world lines of structured bodies. 
Measuring such deviations, due to the existence of cumulative effects, should be favorite with respect to measuring the amplitude of the gravitational wave itself.
\end{abstract}

\section{Introduction}

The equations of motion for an extended body in a given gravitational background were deduced
by Dixon \cite{dixon64,dixon69,dixon70,dixon73,dixon74} (hereafter the \lq\lq relativistic extended body model,"  or simply \lq\lq Dixon's model")  in multipole approximation to any order. 
In the quadrupole approximation they read
\begin{eqnarray}
\label{papcoreqs1}
\frac{DP^{\mu}}{\rmd \tau_U}&=&-\frac12R^{\mu}{}_{\nu\alpha\beta}U^{\nu}S^{\alpha\beta}-\frac16J^{\alpha\beta\gamma\delta}R_{\alpha\beta\gamma\delta}{}^{;\,\mu}
\equiv F^{\rm (spin)}{}^{\mu}+F^{\rm (quad)}{}^{\mu}\ , \\
\label{papcoreqs2}
\frac{DS^{\mu\nu}}{\rmd \tau_U}&=&2P^{[\mu}U^{\nu]}+\frac43J^{\alpha\beta\gamma[\mu}R^{\nu]}{}_{\gamma\alpha\beta}\ ,
\end{eqnarray}
where $P^{\mu}=m U_p^\mu$ (with $U_p\cdot U_p=-1$) is the total four-momentum of the particle, and $S^{\mu\nu}$ is a (antisymmetric) spin tensor; 
$U$ is the timelike unit tangent vector of the \lq\lq center of mass line'' ${\mathcal C}_U$ used to make the multipole reduction, parametrized by the proper time $\tau_U$.
The tensor $J^{\alpha\beta\gamma\delta}$ is the quadrupole moment of the stress-energy tensor of the body, and has the same algebraic symmetries as the Riemann tensor. 
In this paper we limit our considerations to Dixon's model under the further simplifying assumption \cite{taub64,ehlers77} that the only contribution to the complete quadrupole moment $J^{\alpha\beta\gamma\delta}$ stems from the mass quadrupole moment $Q^{\alpha\beta}$, i.e. we write
\beq
\label{Jdef}
J^{\alpha\beta\gamma\delta}=-3U_p^{[\alpha}Q^{\beta][\gamma}U_p^{\delta]}\ ,\qquad Q^{\alpha\beta}U_p{}_\beta=0\ .
\eeq

In order the model to be mathematically consistent the following additional condition should be imposed \cite{dixon64} to the spin tensor
\beq
\label{Tconds}
S^{\mu\nu}U_p{}_\nu=0\ ,
\eeq
to ensure the correct definition of the various multipolar terms.

It is also convenient to introduce the  spin vector by spatial (with respect to $U_p$) duality
\beq
\label{spinvec}
S^\beta={\textstyle\frac12} \eta_\alpha{}^{\beta\gamma\delta}U_p^\alpha S_{\gamma\delta}\ ,
\eeq
where $\eta_{\alpha\beta\gamma\delta}=\sqrt{-g} \epsilon_{\alpha\beta\gamma\delta}$ is the unit volume 4-form and $\epsilon_{\alpha\beta\gamma\delta}$ ($\epsilon_{0123}=1$) is the Levi-Civita alternating symbol, 
 as well as the scalar invariant
\beq
\label{sinv}
s^2=\frac12 S_{\mu\nu}S^{\mu\nu}=S_\mu S^\mu\ , 
\eeq
which is in general not constant along the path.

There are no evolution equations for the quadrupole as well as higher multipoles as a consequence of the Dixon's construction, so their evolution is completely free, depending only on the considered body.
Therefore the system of equations is not self-consistent, and one must assume that all unspecified quantities are  known as intrinsic properties of the matter under consideration.
Moreover, the assumption that the considered body is a test body means that its mass, its spin as well as its quadrupole moments must all be small enough not to contribute significantly to the background metric. Otherwise, backreaction must be taken into account.

We investigate here how a small extended body at rest reacts to the passage of a single plane gravitational wave (GPW) in the weak field limit. The body is spinning and also endowed with a quadrupolar structure. The latter property allows us to consider this analysis as a good model for a bar antenna. 
The present work thus extends previous results by Mohseni and Sepangi \cite{mohseni} concerning the motion of purely spinning particles (i.e. with negligible quadrupolar structure) in a GPW background.

\section{Motion of extended bodies in GPW spacetimes}

Consider the metric of a single gravitational plane wave propagating along the $x$ direction of a coordinate frame, with \lq\lq$+,\times$" polarization states, written in the form
\beq
\label{gwmetric}
\rmd s^2 =-\rmd t^2+\rmd x^2 +(1-h_+)\rmd y^2 +(1+h_+)\rmd z^2-2h_\times\rmd x\rmd z\ ,
\eeq
where $h_{+,\times}=h_{+,\times}(t-x)$ are functions only of $t-x$ because of vacuum linearized Einstein equations. 
Let the metric functions have the following form 
\beq
h_+=A_+\sin\omega (t-x)\ , \qquad
h_\times=A_\times\cos\omega (t-x)\ .
\eeq
A wave with linear polarization is characterized by $A_+=0$ or $A_\times=0$, whereas the condition $A_+=\pm A_\times$ corresponds to a circularly polarized wave. 
It is also useful to introduce the \lq\lq polarization angle'' of the wave $\tan\psi=A_\times/A_+$.
Throughout the papers all tensorial quantities are assumed to be decomposed along the coordinate frame $\{e_\alpha\}\equiv\{\partial_\alpha\}$.

The geodesic of this metric are given by \cite{gyroGW}
\begin{eqnarray}
U_{(\rm geo)}&=&\frac{1}{2E}[(\mu^2+f+E^2)\partial_t+(\mu^2+f-E^2)\partial_x]\nonumber\\
&&+[\alpha(1+h_+)+\beta h_\times]\partial_y+[\beta(1-h_+)+\alpha h_\times]\partial_z\ ,
\end{eqnarray}
where $\alpha$, $\beta$ and $E$ are conserved Killing quantities, $\mu^2=1,0,-1$ corresponding to timelike, null and spacelike geodesics respectively, and
\beq
f=\alpha^2(1+h_+)+\beta^2(1-h_+) +2\alpha\beta h_\times\ .
\eeq
The corresponding parametric equations of the geodesic orbits are then easily obtained
\begin{eqnarray}
t(\lambda)&=&E\lambda+t_0+x(\lambda)-x_0\ , \nonumber\\
x(\lambda)&=&(\mu^2+\alpha^2+\beta^2-E^2)\frac{\lambda}{2E}-\frac{1}{2\omega E^2}[(\alpha^2-\beta^2)A_+\cos\omega(E\lambda+t_0-x_0)\nonumber\\
&&-2\alpha\beta A_\times\sin\omega(E\lambda+t_0-x_0)]+x_0\ , \nonumber\\
y(\lambda)&=&\alpha\lambda+y_0\nonumber\\
&&-\frac{1}{\omega E}[\alpha A_+\cos\omega(E\lambda+t_0-x_0)-\beta A_\times\sin\omega(E\lambda+t_0-x_0)]\ , \nonumber\\
z(\lambda)&=&\beta\lambda+z_0\nonumber\\
&&+\frac{1}{\omega E}[\beta A_+\cos\omega(E\lambda+t_0-x_0)+\alpha A_\times\sin\omega(E\lambda+t_0-x_0)]\ , 
\end{eqnarray}
where $\lambda$ is an affine parameter and $x^\alpha_0$ are integration constants.

For later convenience we also evaluate the world function, which is defined for any given background metric as half the square of the spacetime distance between two generic points $x_A$ and $x_B$ connected by a geodesic path \cite{synge} 
\beq
\Omega(x_A,x_B)=\frac12\int_0^1g_{\mu\nu}[x^\alpha(\lambda)]\frac{\rmd x^\mu}{\rmd\lambda}\frac{\rmd x^\nu}{\rmd\lambda}\rmd\lambda
\equiv\Omega_{AB}\ ,
\eeq
where $x^\alpha(\lambda)$ satisfy the geodesic equations and the affine parameter is such that $x^\alpha(0)=x^\alpha_A$ and $x^\alpha(1)=x^\alpha_B$. 
We obtain
\begin{eqnarray}
\label{worldfun}
\Omega_{AB}&=&\Omega_{AB}^{\rm flat}\nonumber\\
&&+\frac{A_+}{2\omega}[(y_B-y_A)^2-(z_B-z_A)^2]\frac{\cos\omega(t_B-x_B)-\cos\omega(t_A-x_A)}{t_B-x_B-(t_A-x_A)}\nonumber\\
&&-\frac{A_\times}{\omega}(y_B-y_A)(z_B-z_A)\frac{\sin\omega(t_B-x_B)-\sin\omega(t_A-x_A)}{t_B-x_B-(t_A-x_A)}\ ,
\end{eqnarray}
where 
\beq
\Omega_{AB}^{\rm flat}=\frac12\eta_{\alpha\beta}(x_A-x_B)^\alpha(x_A-x_B)^\beta\ 
\eeq
is the well known expression of the world function in the Minkowski spacetime. To our knowledge the expression  (\ref{worldfun}) for the world function cannot be found in the literature.

It is in general a hard task to solve the whole set of equations (\ref{papcoreqs1})--(\ref{Tconds}) in its complete generality, without imposing convenient restrictions on the form of the various quantities involved, e.g. constant frame components of the spin and quadrupole tensors, center of mass line moving along a given orbit, total four-momentum of the body aligned with a given direction \cite{bfgo}.  
However, since we are interested in a weak field solution it is sufficient to search for changes in the mass parameter $m$ of the body, its velocity and total four-momentum as well as spin and quadrupole tensors which are of the same order as the the metric functions $h_{+,\times}$:
\begin{eqnarray}
\label{vincexp}
&&m=m_0+\tilde m\ , \quad U=U_0+\tilde U\ , \quad P=P_0+\tilde P\ , \nonumber\\
&&S^{\mu\nu}=S^{\mu\nu}_0+\tilde S^{\mu\nu}\ , \quad Q^{\mu\nu}=Q^{\mu\nu}_0+\tilde Q^{\mu\nu}\ ,
\end{eqnarray}
where the quantities with the subscript \lq\lq0'' are the flat background ones, i.e. those corresponding to the initial values, before the passage of the wave. 
By taking advantage from that we are able to solve the problem in its most general form. 
The background quantities are solutions of the zeroth-order set of equations 
\beq
\label{flateqs}
\frac{DP_0^{\mu}}{\rmd \tau_{U_0}}=0\ , \qquad \frac{DS^{\mu\nu}_0}{\rmd \tau_{U_0}}=2P_0^{[\mu}U_0^{\nu]}\ .
\eeq
Assume that the body is initially at rest at the origin of the coordinates, i.e. the associated world line has parametric equations
\beq
\label{traietiniz}
t=\tau_{U_0}\ , \quad x(\tau_{U_0})=0\ , \quad y(\tau_{U_0})=0\ , \quad z(\tau_{U_0})=0\ ,
\eeq
where $\tau_{U_0}$ denotes the proper time parameter, and hence the unit tangent vector reduces to $U_0=\partial_t$.
The set of equations (\ref{flateqs}) is fulfilled for instance by taking $P_0=m_0U_0$, implying that the components of the background spin tensor $S^{\mu\nu}_0$ remain constant along the path.
The conditions (\ref{Tconds}) simply give $S^{0a}_0=0$. The remaining components are related to the background spin vector components by 
\beq
S^{12}_0=S^3_0\ , \qquad S^{13}_0=-S^2_0\ , \qquad S^{23}_0=S^1_0\ .
\eeq
Note that the quadrupolar quantities do not enter the flat spacetime equations, since they are coupled to the Riemann tensor. Therefore, we can assume the components of the background quadrupole tensor $Q^{\mu\nu}_0$ to be also constant with respect to the coordinate frame, without loss in generality.
The conditions (\ref{Jdef}) give $Q^{0\alpha}_0=0$.

The conditions ensuring that the body does not perturb the metric can be formulated by using the natural length scales associated with it, i.e. the \lq\lq bare'' mass $m_0$, $|S_0^\alpha|/m_0$ and $(|Q_0^{\alpha\beta}|/m_0)^{1/2}$.

Let us turn to the general system of equations (\ref{papcoreqs1})--(\ref{Tconds}). 
The first order set of equations is obtained after substituting in it the perturbed quantities (\ref{vincexp}), by retaining terms up to the first order in the metric functions.
Let the \lq\lq center of mass line'' be generic after the passage of the wave, i.e. with timelike unit tangent vector $U$ given by 
\beq
U=\gamma(\partial_t+\tilde \nu^a \partial_a)\ , \quad \gamma=(1-\tilde \nu^2)^{-1/2}\ , \quad \tilde \nu^2=\delta_{ab}\tilde \nu^a\tilde \nu^b\ , \quad a,b=1,2,3\ .
\eeq
According to the decomposition (\ref{vincexp}) the linear velocities along the spatial axes are first order in $h_{+,\times}$, so that $\gamma\simeq1$.
Analogously, let the total four-momentum $P=mU_p$ be generic, with 
\beq
U_p=\gamma_p(\partial_t+\tilde \nu_p^a \partial_a)\ , \qquad \gamma_p=(1-\tilde \nu_p^2)^{-1/2}\ , \qquad \tilde \nu_p^2=\delta_{ab}\tilde \nu_p^a\tilde \nu_p^b\ ;
\eeq
also in this case $\gamma_p\simeq1$.
The supplementary conditions (\ref{Tconds}) imply that 
\beq
\label{Tconds2}
\tilde S^{01}=S^3_0\tilde \nu_p^2-S^2_0\tilde \nu_p^3\ , \quad 
\tilde S^{02}=-S^3_0\tilde \nu_p^1+S^1_0\tilde \nu_p^3\ , \quad 
\tilde S^{03}=S^2_0\tilde \nu_p^1-S^1_0\tilde \nu_p^2\ .
\eeq
The spin vector is orthogonal to $U_p$ from its definition (\ref{spinvec}), with components
\begin{eqnarray}
\label{spinveccompts}
&S^0=S_0^1\tilde \nu^1+S_0^2\tilde \nu^2+S_0^3\tilde \nu^3\ , &\quad 
S^1=S_0^1+\tilde S^{23}\ ,  \nonumber\\
&S^2=S_0^2-\tilde S^{13}\ , &\quad 
S^3=S_0^3+\tilde S^{12}\ .	
\end{eqnarray}
Finally, the conditions (\ref{Jdef})$_2$ on the components of the first order quadrupole tensor give
\beq
\label{Qconds}
\tilde Q^{00}=0\ , \quad \tilde Q^{0a}=-Q^{1a}_0\tilde \nu_p^1-Q^{2a}_0\tilde \nu_p^2-Q^{3a}_0\tilde \nu_p^3\ .
\eeq 

It is useful to introduce the rescaled dimensionless quantities (for both zeroth order and first order terms)
\beq
\label{sigmaandq}
\sigma^\alpha=\frac{S^\alpha}{m_0}\omega\ , \qquad q^{\alpha\beta}=\frac{Q^{\alpha\beta}}{m_0}\omega^2\ .
\eeq
Note that the components of the zeroth order quadrupole tensor all have definite sign, whereas the components of the spin vector are allowed to have both signs, corresponding to a body spinning up/down.

Consider first the equations of motion (\ref{papcoreqs1}). 
The spin force and the quadrupole force as defined in Eq. (\ref{papcoreqs1}) turn out to act on the orthogonal two-planes $y-z$ and $t-x$ respectively:
\begin{eqnarray}
F^{\rm (spin)}&=&\frac{m_0\omega}{2}\left[(h_+\sigma^3_0-h_\times \sigma^2_0)\partial_y+(h_\times \sigma^3_0+h_+ \sigma^2_0)\partial_z\right]\ , \nonumber\\ 
F^{\rm (quad)}&=&\frac{m_0\omega}{2}\left[\cot\psi h_\times f_0-2\tan\psi h_+q^{23}_0\right](\partial_t-\partial_x)\ ,
\end{eqnarray}
where we have introduced the quantity $f_0=q^{22}_0-q^{33}_0$.
Eqs. (\ref{papcoreqs1}) thus reduce to the following set
\begin{eqnarray}
-\frac{\rmd \tilde m}{\rmd \tau_U}&=F^{\rm (quad)}{}^0\ , \qquad &
m_0\frac{\rmd \tilde \nu_p^1}{\rmd \tau_U}=F^{\rm (quad)}{}^1\ , \nonumber\\
m_0\frac{\rmd \tilde \nu_p^2}{\rmd \tau_U}&=F^{\rm (spin)}{}^2\ , \qquad &
m_0\frac{\rmd \tilde \nu_p^3}{\rmd \tau_U}=F^{\rm (spin)}{}^3\ ,
\end{eqnarray}
whose solution is straightforward:
\begin{eqnarray}
\label{solmoto1}
\tilde m&=&-\frac{m_0}{4}\left[f_0A_+\sin\omega\tau_U+2q^{23}_0A_\times(\cos\omega\tau_U-1)\right]\ , \nonumber\\
\tilde \nu_p^1&=&\frac{\tilde m}{m_0}-\frac{A_\times}{2}q^{23}_0+c_1
=-\frac{1}{4}\left[f_0A_+\sin\omega\tau_U+2q^{23}_0A_\times\cos\omega\tau_U\right]+c_1\ , \nonumber\\
\tilde \nu_p^2&=&-\frac12\left[\sigma^2_0A_\times\sin\omega\tau_U+\sigma^3_0A_+\cos\omega\tau_U\right]+c_2\ , \nonumber\\
\tilde \nu_p^3&=&\frac12\left[\sigma^3_0A_\times\sin\omega\tau_U-\sigma^2_0A_+\cos\omega\tau_U\right]+c_3\ ,
\end{eqnarray}
where the initial condition $\tilde m(0)=0$ has been imposed, and $c_1$, $c_2$ and $c_3$ are integration constants.
Eq. (\ref{solmoto1})$_1$ implies that the zeroth order quadrupole tensor is responsible for the mass of the body to vary after the passage of the wave. 

Consider then the evolution equations (\ref{papcoreqs2}) for the spin tensor.
By using the supplementary conditions (\ref{Tconds2}) they give three algebraic relation between the spatial linear velocities $\tilde \nu^a$ of $U$ and $\tilde \nu_p^a$ of $U_p$ plus three evolution equations for the spatial components $\tilde S^a$ of the spin tensor to be integrated together with the initial conditions $\tilde \sigma^a(0)=0$.
The corresponding solution turns out to be given by
\begin{eqnarray}
\label{solnuandS}
\tilde \nu^1&=&\frac14[2((\sigma^2_0)^2-(\sigma^3_0)^2)+f_0]A_+\sin\omega\tau_U + \frac12(2\sigma^2_0\sigma^3_0+q^{23}_0)A_\times(\cos\omega\tau_U-1)\ , \nonumber\\
\tilde \nu^2&=&-\frac12\left[A_+(\sigma^1_0\sigma^2_0+q^{12}_0)+A_\times(\sigma^2_0-\sigma^3_0q^{23}_0)\right]\sin\omega\tau_U\nonumber\\
&&-\frac14\left[2A_\times(\sigma^1_0\sigma^3_0+q^{13}_0)+A_+\sigma^3_0(2+f_0)\right](\cos\omega\tau_U-1)\ , \nonumber\\
\tilde \nu^3&=&\frac12\left[A_+(\sigma^1_0\sigma^3_0+q^{13}_0)+A_\times(\sigma^3_0-\sigma^2_0q^{23}_0)\right]\sin\omega\tau_U\nonumber\\
&&-\frac14\left[2A_\times(\sigma^1_0\sigma^2_0+q^{12}_0)+A_+\sigma^2_0(2-f_0)\right](\cos\omega\tau_U-1)\ , \nonumber\\
\tilde \sigma^1&=&\frac12f_0A_\times\sin\omega\tau_U+q^{23}_0A_+(\cos\omega\tau_U-1)\ , \nonumber\\
\tilde \sigma^2&=&\frac12(A_\times \sigma^3_0-q^{13}_0A_+)(\cos\omega\tau_U-1)-\frac12(A_+ \sigma^2_0-q^{12}_0A_\times)\sin\omega\tau_U\ , \nonumber\\
\tilde \sigma^3&=&\frac12(A_\times \sigma^2_0-q^{12}_0A_+)(\cos\omega\tau_U-1)-\frac12(A_+ \sigma^3_0-q^{13}_0A_\times)\sin\omega\tau_U\ ,
\end{eqnarray}
where the constant $c_1$, $c_2$ and $c_3$ have been conveniently fixed so that $\tilde \nu^a(0)=0$, implying that
\begin{eqnarray}
\label{solmoto2}
\tilde \nu_p^1&=&-\frac{1}{4}\left[f_0A_+\sin\omega\tau_U+2q^{23}_0A_\times(\cos\omega\tau_U-1)\right]-(\sigma^2_0\sigma^3_0+q^{23}_0)A_\times\ , \nonumber\\
\tilde \nu_p^2&=&-\frac12\left[\sigma^2_0A_\times\sin\omega\tau_U+\sigma^3_0A_+(\cos\omega\tau_U-1)\right]\nonumber\\
&&+\frac14\left[2A_\times(\sigma^1_0\sigma^3_0+q^{13}_0)+A_+\sigma^3_0f_0\right]\ , \nonumber\\
\tilde \nu_p^3&=&\frac12\left[\sigma^3_0A_\times\sin\omega\tau_U-\sigma^2_0A_+(\cos\omega\tau_U-1)\right]\nonumber\\
&&+\frac14\left[2A_\times(\sigma^1_0\sigma^2_0+q^{12}_0)-A_+\sigma^2_0f_0\right]\ .
\end{eqnarray} 
Eqs. (\ref{solnuandS})$_{4,5,6}$ thus imply that even if initially absent the spinning structure will be acquired by the body during the evolution due to its quadrupolar structure.

Finally, the nonvanishing components of the first order quadrupole tensor can be easily obtained from Eqs. (\ref{Qconds}), whereas $\tilde \sigma^0$ is given by Eq. (\ref{spinveccompts}).

The modification to the initial trajectory (\ref{traietiniz}) of the body after the passage of the wave is easily obtained by integrating Eqs. (\ref{solnuandS})$_{1,2,3}$, since $\tilde \nu^a={\rmd x^a}/{\rmd \tau_U}$, together with the initial conditions $x^a(0)=0$:
\begin{eqnarray}
\label{traietfin}	
t&=&\tau_U\ , \nonumber\\
\omega\, x&=&-\frac14[2((\sigma^2_0)^2-(\sigma^3_0)^2)+f_0]A_+(\cos\omega\tau_U-1)\nonumber\\
&&+\frac12[2\sigma^2_0\sigma^3_0+q^{23}_0]A_\times(\sin\omega\tau_U-\omega\tau_U)\ , \nonumber\\
\omega\, y&=&\frac12\left[A_+(\sigma^1_0\sigma^2_0+q^{12}_0)+A_\times(\sigma^2_0-\sigma^3_0q^{23}_0)\right](\cos\omega\tau_U-1)\nonumber\\
&&-\frac14\left[2A_\times(\sigma^1_0\sigma^3_0+q^{13}_0)+A_+\sigma^3_0(2+f_0)\right](\sin\omega\tau_U-\omega\tau_U)\ , \nonumber\\
\omega\, z&=&-\frac12\left[A_+(\sigma^1_0\sigma^3_0+q^{13}_0)+A_\times(\sigma^3_0-\sigma^2_0q^{23}_0)\right](\cos\omega\tau_U-1)\nonumber\\
&&-\frac14\left[2A_\times(\sigma^1_0\sigma^2_0+q^{12}_0)+A_+\sigma^2_0(2-f_0)\right](\sin\omega\tau_U-\omega\tau_U)\ . 
\end{eqnarray}

\section{Discussion and general results}

In the expressions (\ref{traietfin}) we can distinguish a harmonic part and a secular one; the latter is given by
\begin{eqnarray}	
\label{secular}
\omega\, x_{\rm sec}&=&-\frac{A_\times}2(2\sigma^2_0\sigma^3_0+q^{23}_0)\,\omega\tau_U\ , \nonumber\\
\omega\, y_{\rm sec}&=&\frac14\left[2A_\times(\sigma^1_0\sigma^3_0+q^{13}_0)+A_+\sigma^3_0(2+f_0)\right]\omega\tau_U\ , \nonumber\\
\omega\, z_{\rm sec}&=&\frac14\left[2A_\times(\sigma^1_0\sigma^2_0+q^{12}_0)+A_+\sigma^2_0(2-f_0)\right]\omega\tau_U\ , 
\end{eqnarray} 
and it is important in view of \lq\lq cumulative'' effects which it originates.

It is now interesting to evaluate the deviations between the geodesic world line of a reference test particle (always) at rest at the origin of the coordinates and that of the structured body under consideration here, initially (only) at rest at the origin of the coordinates, i.e. with unit tangent vector  initially aligned with the geodesic one.
The deviations can be obtained by evaluating the world function (\ref{worldfun}) at the same coordinate time $\tau_{U_0}=t=\tau_U$, where $x_A^\alpha=x^\alpha|_{\tau_{U_0}}$ and $x_B^\alpha=x^\alpha|_{\tau_{U}}$ are given by Eqs. (\ref{traietiniz}) and (\ref{traietfin}), respectively.
The world function is given by
\beq
\Omega_{AB}=\frac12\delta_{ab}x^ax^b\ ,
\eeq 
by neglecting higher order terms in the amplitudes $A_+$ and $A_\times$ of the gravitational wave.
Apart from oscillating contributions, the dominant one is 
\beq
\Omega_{AB}\simeq\frac12[x_{\rm sec}^2+y_{\rm sec}^2+z_{\rm sec}^2]\ ,
\eeq 
so that the deviation turns out to be
\beq
\label{deviation}
d_{AB}\equiv\sqrt{2|\Omega_{AB}|}\simeq\sqrt{x_{\rm sec}^2+y_{\rm sec}^2+z_{\rm sec}^2}\ .
\eeq 
Moreover, assuming $A_+\sim A_\times\sim h$, from Eqs. (\ref{secular}) it follows that the deviation $d_{AB}\propto h \tau_U$, the constant of proportionality depending on the proper structure of the body.  It is therefore clear that cumulative effects arise and can be used to get 
better conditions to measure (indirectly) the amplitude of  a gravitational wave.

The complementary cases of an initially non-spinning body with a given quadrupole structure as well as that of a purely spinning body with a negligible quadrupole structure will be considered in detail in the Appendix. The results concerning two simple cases are listed below.

Consider first the case of an initially axially symmetric (about the $z$-axis) non-spinning body described by the zeroth order quadrupole tensor
\beq
\label{qassiale}
q_0^{ab}={\rm diag}\,[f_0/3,f_0/3,-2f_0/3]\ ,
\eeq
where the tracefree property has been assumed in analogy with the classical (non-relativistic) case.
Eqs. (\ref{orbitsoloquad}) imply no secular deviation, since
\beq
\label{devsoloquad}
d_{AB}=\frac{h|f_0|}{4\omega}|\cos\omega\tau_U-1|\ .
\eeq 
Note that for a non-spinning body secular deviations can arise only for nonvanishing nondiagonal components of the quadrupole tensor, as from Eqs. (\ref{secular}). 

Consider then the case of a purely spinning body along the $y$-axis for instance, with vanishing quadrupole tensor.
Eqs. (\ref{orbitsolospin}) imply a secular deviation given by
\beq
d_{AB}\simeq \frac{h|\sigma_0^2|}{2}\tau_U\ .
\eeq

\section{Concluding remarks}

We have studied how a small extended body at rest interacts with an incoming single plane gravitational wave. The body is spinning and also endowed with a quadrupolar structure, so that due to the latter property it can be thus considered as a good model for a gravitational wave antenna. 
We have discussed the motion of such an extended body by assuming that it can be described according to Dixon's model and that the gravitational field of the wave is weak, i.e. the \lq\lq reaction'' (induced motion) of a \lq\lq gravitational wave antenna" (the extended body) to the passage of the wave.

We have found that in general, even if initially absent, the body acquires a dipolar moment induced by the quadrupole tensor, a property never pointed out before in the literature.
As a byproduct, the derivation of the \lq\lq world function'' in the weak field limit of a gravitational wave background (absent in the literature to our knowledge) has been accomplished and widely used throughout the paper.   
In particular, we have applied it to estimate the orbital deviations between geodesics and the world lines of structured bodies. 
Such deviations could be measured by gravitational wave detectors.
Furthermore, due to the existence of cumulative effects, measuring the deviation of the center of mass of the body as induced by the passage of a gravitational wave should be favorite with respect to measuring the amplitude of the gravitational wave itself, whose estimate can also be achieved indirectly, since it is simply related to the magnitude of the deviation.

\appendix

\section{Limiting cases}

We will examine separately below the complementary cases of an initially non-spinning body with a given quadrupole structure as well as that of a purely spinning body with a negligible quadrupole structure.

\subsection{Body initially not spinning}

In the special case in which the body is initially not spinning (i.e. $\sigma_0^a=0$), the general solutions (\ref{solmoto1})--(\ref{solmoto2}) reduce to
\begin{eqnarray}
\label{soloquad}
\tilde m&=&-\frac{m_0}{4}\left[f_0A_+\sin\omega\tau_U+2q^{23}_0A_\times(\cos\omega\tau_U-1)\right]\ , \nonumber\\
\tilde \nu_p^1&=&\frac{\tilde m}{m_0}-q^{23}_0A_\times\ , \nonumber\\
\tilde \nu_p^2&=&\frac12q^{13}_0A_\times\ , \nonumber\\
\tilde \nu_p^3&=&\frac12q^{12}_0A_\times\ , \nonumber\\
\tilde \nu^1&=&\frac14f_0A_+\sin\omega\tau_U + \frac12q^{23}_0A_\times(\cos\omega\tau_U-1)\ , \nonumber\\
\tilde \nu^2&=&-\frac12A_+q^{12}_0\sin\omega\tau_U-\frac12A_\times q^{13}_0(\cos\omega\tau_U-1)\ , \nonumber\\
\tilde \nu^3&=&\frac12A_+q^{13}_0\sin\omega\tau_U-\frac12A_\times q^{12}_0(\cos\omega\tau_U-1)\ , \nonumber\\
\tilde \sigma^1&=&\frac12f_0A_\times\sin\omega\tau_U+q^{23}_0A_+(\cos\omega\tau_U-1)\ , \nonumber\\
\tilde \sigma^2&=&-\frac12A_+q^{13}_0(\cos\omega\tau_U-1)+\frac12A_\times q^{12}_0\sin\omega\tau_U\ , \nonumber\\
\tilde \sigma^3&=&-\frac12A_+q^{12}_0(\cos\omega\tau_U-1)+\frac12A_\times q^{13}_0\sin\omega\tau_U\ ,
\end{eqnarray}
and the spatial orbit becomes 
\begin{eqnarray}
\omega\, x&=&-\frac14f_0A_+(\cos\omega\tau_U-1)+\frac12q^{23}_0A_\times(\sin\omega\tau_U-\omega\tau_U)\ , \nonumber\\
\omega\, y&=&\frac12A_+q^{12}_0(\cos\omega\tau_U-1)-\frac12A_\times q^{13}_0(\sin\omega\tau_U-\omega\tau_U)\ , \nonumber\\
\omega\, z&=&-\frac12A_+q^{13}_0(\cos\omega\tau_U-1)-\frac12A_\times q^{12}_0(\sin\omega\tau_U-\omega\tau_U)\ . 
\end{eqnarray}
From Eqs. (\ref{soloquad}) it is evident that even initially non-spinning the body starts to spin.

If the body under consideration is such that $q_0^{\alpha\beta}$ is diagonal we find
\begin{eqnarray}
\tilde m&=&-\frac{m_0}{4}f_0A_+\sin\omega\tau_U\ , \nonumber\\
\tilde \nu_p^1&=&-\frac{1}{4}f_0A_+\sin\omega\tau_U=\frac{\tilde m}{m_0}\ , \quad \tilde \nu_p^2=0\ , \quad \tilde \nu_p^3=0\ , \nonumber\\
\tilde \nu^1&=&\frac14f_0A_+\sin\omega\tau_U=-\frac{\tilde m}{m_0}\ , \quad \tilde \nu^2=0\ , \quad \tilde \nu^3=0\ , \nonumber\\
\tilde \sigma^1&=&\frac12f_0A_\times\sin\omega\tau_U=-2\frac{A_\times}{A_+}\frac{\tilde m}{m_0}\ , \quad \tilde \sigma^2=0\ , \quad \tilde \sigma^3=0\ ,
\end{eqnarray}
and the spatial orbit becomes 
\beq
\label{orbitsoloquad}
\omega\, x=-\frac14f_0A_+(\cos\omega\tau_U-1)\ , \quad 
\omega\, y=0\ , \quad
\omega\, z=0\ . 
\eeq

\subsection{Purely spinning particle}

In the case of a purely spinning particle, i.e. with vanishing components of the quadrupole tensor, we have:
\begin{eqnarray}
\tilde m&=&0\ , \nonumber\\
\tilde \nu_p^1&=&-A_\times\sigma^2_0\sigma^3_0\ , \nonumber\\
\tilde \nu_p^2&=&-\frac12\left[\sigma^2_0A_\times\sin\omega\tau_U+\sigma^3_0A_+(\cos\omega\tau_U-1)\right]+\frac12A_\times\sigma^1_0\sigma^3_0\ , \nonumber\\
\tilde \nu_p^3&=&\frac12\left[\sigma^3_0A_\times\sin\omega\tau_U-\sigma^2_0A_+(\cos\omega\tau_U-1)\right]+\frac12A_\times\sigma^1_0\sigma^2_0\ , \nonumber\\
\tilde \nu^1&=&\frac12[(\sigma^2_0)^2-(\sigma^3_0)^2]A_+\sin\omega\tau_U + \sigma^2_0\sigma^3_0A_\times(\cos\omega\tau_U-1)\ , \nonumber\\
\tilde \nu^2&=&-\frac12\left[A_+\sigma^1_0\sigma^2_0+A_\times\sigma^2_0\right]\sin\omega\tau_U\nonumber\\
&&-\frac12\left[A_\times\sigma^1_0\sigma^3_0+A_+\sigma^3_0\right](\cos\omega\tau_U-1)\ , \nonumber\\
\tilde \nu^3&=&\frac12\left[A_+\sigma^1_0\sigma^3_0+A_\times\sigma^3_0\right]\sin\omega\tau_U\nonumber\\
&&-\frac12\left[A_\times\sigma^1_0\sigma^2_0+A_+\sigma^2_0\right](\cos\omega\tau_U-1)\ , \nonumber\\
\tilde \sigma^1&=&0\ , \nonumber\\
\tilde \sigma^2&=&\frac12A_\times \sigma^3_0(\cos\omega\tau_U-1)-\frac12A_+ \sigma^2_0\sin\omega\tau_U\ , \nonumber\\
\tilde \sigma^3&=&\frac12A_\times \sigma^2_0(\cos\omega\tau_U-1)-\frac12A_+ \sigma^3_0\sin\omega\tau_U\ ,
\end{eqnarray}
and the spatial orbit
\begin{eqnarray}
\omega\, x&=&-\frac12[(\sigma^2_0)^2-(\sigma^3_0)^2]A_+(\cos\omega\tau_U-1)+\sigma^2_0\sigma^3_0A_\times(\sin\omega\tau_U-\omega\tau_U)\ , \nonumber\\
\omega\, y&=&\frac{\sigma^2_0}{2}(A_+\sigma^1_0+A_\times)(\cos\omega\tau_U-1)-\frac{\sigma^3_0}{2}(A_\times\sigma^1_0+A_+)(\sin\omega\tau_U-\omega\tau_U)\ , \nonumber\\
\omega\, z&=&-\frac{\sigma^3_0}{2}(A_+\sigma^1_0+A_\times)(\cos\omega\tau_U-1)-\frac{\sigma^2_0}{2}(A_\times\sigma^1_0+A_+)(\sin\omega\tau_U-\omega\tau_U),\  
\end{eqnarray} 
in agreement with the results of \cite{mohseni}.

In the special case where the initial spin of the particle is aligned with a transverse spatial direction, e.g. $\sigma^1_0=0=\sigma^3_0$, the above relations have the following limit:
\begin{eqnarray}
\tilde m&=&0\ , \nonumber\\
\tilde \nu_p^1&=&0\ , \quad
\tilde \nu_p^2=-\frac12A_\times\sigma^2_0\sin\omega\tau_U\ , \quad
\tilde \nu_p^3=-\frac12A_+\sigma^2_0(\cos\omega\tau_U-1)\ , \nonumber\\
\tilde \nu^1&=&\frac12A_+(\sigma^2_0)^2\sin\omega\tau_U\ , \,
\tilde \nu^2=-\frac12A_\times\sigma^2_0\sin\omega\tau_U\ , \,
\tilde \nu^3=-\frac12A_+\sigma^2_0(\cos\omega\tau_U-1)\ , \nonumber\\
\tilde \sigma^1&=&0\ , \quad
\tilde \sigma^2=-\frac12A_+ \sigma^2_0\sin\omega\tau_U\ , \quad
\tilde \sigma^3=\frac12A_\times \sigma^2_0(\cos\omega\tau_U-1)\ ,
\end{eqnarray}
and the spatial orbit
\begin{eqnarray}
\label{orbitsolospin}
\omega\, x&=&-\frac12A_+(\sigma^2_0)^2(\cos\omega\tau_U-1)\ , \nonumber\\
\omega\, y&=&\frac12A_\times\sigma^2_0(\cos\omega\tau_U-1)\ , \nonumber\\
\omega\, z&=&-\frac12A_+\sigma^2_0(\sin\omega\tau_U-\omega\tau_U)\ .
\end{eqnarray}

\end{document}